# Strong Chaos without Butterfly Effect in Dynamical Systems with Feedback

Guido Boffetta[1], Giovanni Paladin[2], and Angelo Vulpiani[3]

[1] *Istituto di Fisica Generale, Università di Torino*
*Via P.Giuria 1 I-10125 Torino, Italy*

[2] *Dipartimento di Fisica, Università dell'Aquila*
*Via Vetoio I-67100 Coppito, L'Aquila, Italy*

[3] *Dipartimento di Fisica, Università di Roma 'La Sapienza'*
*P.le A.Moro 2 I-00185 Roma, Italy*

ABSTRACT

We discuss the predictability of a conservative system that drives a chaotic system with positive maximum Lyapunov exponent $\lambda_0$, such as the erratic motion of an asteroid in the gravitational field of two bodies of much larger mass. We consider the case where in absence of feedback (restricted model), the driving system is regular and completely predictable. A small feedback of strength $\epsilon$ still allows a good forecasting in the driving system up to a very long time $T_p \sim \epsilon^{-\alpha}$, where $\alpha$ depends on the details of the system. The most interesting situation happens when the Lyapunov exponent of the total system is strongly chaotic with $\lambda_{tot} \approx \lambda_0$, practically independent of $\epsilon$. Therefore an exponential amplification of a small incertitude on the initial conditions in the driving system for any $\epsilon \neq 0$ coexists with very long predictability times. The paradox stems from saturation effects in the evolution for the growth of the incertitude as illustrated in a simple model of coupled maps and in a system of three point vortices in a disk.

It is commonly believed that a sensible dependence on initial condition makes forecasting impossible even in systems with few degrees of freedom. This is the so-called butterfly effect discovered by Lorenz in a numerical simulation of a model of convection with three degrees of freedom [1]. Using his words, 'A butterfly moving its wings over Brazil might cause the formation of a tornado in Texas'.

In general, a dynamical system is considered chaotic when there is an exponential amplification of an infinitesimal perturbation $\delta_0$ on the initial conditions, with a mean time rate given by the inverse of the maximum Lyapunov exponent $\lambda$ [2]. Indeed, such a deterministic system is expected to be predictable on times $t \ll \lambda^{-1}$ and to behave like a random system on larger times. The purpose of this letter is to show that there exists a wide class of dynamical systems where a large value of the Lyapunov exponent does not imply a short predictability time on a physically relevant part of the system. In this respect, one can speak of strong chaos $\lambda \gg 0$ without butterfly effect.

In particular, we discuss the predictability of a conservative system that drives a strongly chaotic system with positive maximum Lyapunov exponent $\lambda_0$. In absence of feedback the driving system is regular and completely predictable. A small feedback of strength $\epsilon$ still allows to predict the future of the driving system up to a very long predictability time $T_p$ that diverges with $\epsilon$. The Lyapunov exponent of the total system is $\lambda_{tot} \approx \lambda_0$, and so there is a regime of strong chaos for all $\epsilon$ values. The absence of the butterfly effect stems from saturation effects in the evolution laws for the growth of an incertitude on the driven system

To be explicit, let us consider a system with evolution given by two sets of equations

$$\frac{d\xi}{dt} = F(\xi) + \epsilon\, h(\eta) \tag{1a}$$

$$\frac{d\eta}{dt} = G(\xi, \eta) \tag{1b}$$

with $\xi \in R^n$ and $\eta \in R^m$. The variables $\eta$ are thus driven by a sub-system represented by the variables $\xi$, with a weak feedback of order $\epsilon$. A typical physical example is given by an asteroid moving in the gravitational field generated by two celestial bodies of much larger mass such as Jupiter and Sun [3]. Usually the feedback is neglected, and one considers the restricted three body problem, i.e. the $\eta$ variables passively driven. However, a finer description should take into account even the influence of the asteroid on the evolution of the other two bodies, i.e. an 'active' driving where $\epsilon \neq 0$. This situation appears in many other phenomena, such as the active advection of a contaminant in a fluid, a simple example which will be discussed in this letter.

The main properties of the system in absence of feedback, are the following:
(1) the driver is an independent dynamical system that exhibits a regular evolution with zero Lyapunov exponent;
(2) the driven sub-system is chaotic with positive maximum Lyapunov exponent, say $\lambda_0$.

In other terms, the behavior of the driver ($\xi$ variables) is completely predictable. However, as soon as $\epsilon \neq 0$, one should consider the total system which is obviously chaotic with a Lyapunov exponent $\lambda_{tot}$ that, a part small correction of order $\epsilon$, is given by the Lyapunov exponent $\lambda_0$ of the chaotic driven sub-system. This means that there is an exponential amplification of a small incertitude on the knowledge of the initial conditions even in the driver. This is an amazing result, since it is natural to expect that it is possible to forecast the behavior of the driver for very long times as $\epsilon \to 0$. Actually, intuition is correct while the Lyapunov analysis gives completely wrong hints on the predictability problem at difference with what commonly believed. The famous butterfly effect of Lorenz seems not to forbid the possibility of predicting the future of a part of the system. The paradox stems from saturation effects in the evolution for the growth of the incertitude. To fix notation and definitions, let us consider the evolution of the total system

$$\frac{dx_i}{dt} = f_i(\mathbf{x}) \qquad \text{where } \mathbf{x} = (\xi, \eta) \in R^{n+m} \tag{2}$$

The incertitude on its state is $\Delta(t) = \mathbf{x}(t) - \mathbf{x}'(t)$ where $\mathbf{x}$ and $\mathbf{x}'$ are trajectories starting from close initial conditions, i.e. $|\mathbf{x}(0) - \mathbf{x}'(0)| = \delta_0$. In the limit $\delta_0 \to 0$, $\Delta$ can be confused with the the tangent vector $\mathbf{z}$ whose evolution equations are

$$\frac{d\mathbf{z}}{dt} = J(t)\,\mathbf{z} \qquad \text{where } J_{ik}(t) = \left.\frac{df_i}{dx_j}\right|_{\mathbf{x}(t)} \tag{3}$$

The maximum Lyapunov exponent is then defined as the exponential rate of the incertitude growth,

$$\lambda = \lim_{t\to\infty}\lim_{\delta_0\to 0} \frac{1}{t} \ln\left(\frac{|\Delta(t)|}{\delta_0}\right) = \lim_{t\to\infty} \frac{1}{t} \ln|\mathbf{z}(t)| \tag{4}$$

It is worth stressing that the full equations for the evolution of an incertitude are non-linear:

$$\frac{d\Delta}{dt} = J(t)\,\Delta + O(\Delta^2) \tag{5}$$

so that the two limits in (4) cannot be interchanged.

The predictability of the system is defined in terms of the allowed maximal ignorance on the state of the system, a tolerance parameter $\Delta_{max}$ which must be fixed according the necessities of the observer. The predictability time is thus

$$T_p = \sup_t \{t \text{ such that}|\Delta(t')| \leq \Delta_{max} \text{ for } t' \leq t\} \tag{6}$$

If $\Delta_{max} \ll 1$, (5) is well approximated by (3) and the predictability time can be roughly identified with the inverse Lyapunov exponent, since

$$T_p \sim \frac{1}{\lambda} \ln\left(\frac{\Delta_{max}}{\delta_0}\right) \tag{7}$$

and the dependence on the initial error $\delta_0$ and on the tolerance parameter $\Delta_{max}$ is only logarithmic and can be safely ignored for many practical purposes.

Suppose now to be interested only on the incertitude $\Delta^{(\xi)}$ in the driver system. When $\epsilon = 0$, the Lyapunov exponent of the driver $\lambda_\xi = 0$ so that it is fully predictable, (we have in general $T_p^{(\xi)} \sim \delta_0^{-\beta}$, the exponent $\beta$ depending on the particular system), while the driven system has a Lyapunov exponent $\lambda_0 > 0$. However, for any $\epsilon \neq 0$ the two sub-systems are coupled, and the global Lyapunov exponent is expected to be

$$\lambda = \lambda_0 + O(\epsilon) \tag{8}$$

A direct application of (7) would give

$$T_p^{(\xi)} \sim T_p \sim \frac{1}{\lambda_0} \tag{9}$$

One thus obtains a singular limit, $\lim_{\epsilon \to 0} T_p^{(\xi)}(\epsilon) \neq T_p^{(\xi)}(\epsilon = 0)$. The troubles stem from the identification between incertitude $\Delta(t)$ and tangent vector $\mathbf{z}(t)$, which is not correct on long time scales.

It is convenient to illustrate the problem in a simpler context as the main qualitative aspects of equations (1a) and (1b) can be reproduced considering only the feedback effect in two coupled maps of the type

$$\xi_{t+1} = L \, \xi_t + \epsilon \, h(\eta_t) \tag{10a}$$

$$\eta_{t+1} = G(\eta_t) \tag{10b}$$

where the time $t$ is an integer variable, $\xi \in R^2$, $h = (h_1, h_2)$ is a vector, function of the variable $\eta \in R^1$ whose evolution is ruled by a chaotic one-dimensional map $G$, and $L$ is the linear operator corresponding to a rotation of an arbitrary angle $\theta$,

$$L = \begin{pmatrix} cos(\theta) & -sin(\theta) \\ sin(\theta) & cos(\theta) \end{pmatrix} .$$

When $\epsilon = 0$, one is left with two independent systems, one of them regular and of Hamiltonian type, the other fully chaotic.

These maps provide a simple, maybe the simplest, example of a system with two different temporal regimes:
(A) short times where $\delta_0 exp(\lambda_0 \, t) \ll 1$ so that it is correct to ignore the nonlinear term in (5), so that $\Delta \sim \mathbf{z}$
(B) long times where one should consider the full non-linear equation (5) for the incertitude growth.

From the observer point of view, both these regimes might be interesting, according his particular exigence. If one is interested in forecasting the very fine details of the systems, the tolerance threshold $\Delta_{max}$ could be quite small, hence $T_p \sim \lambda_0^{-1}$. In general, however, a system is considered unpredictable when the incertitude is rather large (say discrimination between sun/rain in meteorology) and regime (B) is the relevant one. In that case, non-linear effects in (5) cannot be neglected and in order to give an analytic estimate of the predictability time we can use a stochastic model of the deterministic equations. Indeed the chaotic feedback on the evolution of the 'driver' system can be simulated by a random vector $w$, i.e.

$$\xi_{t+1} = L \, \xi_t + \epsilon \, w_t \tag{11}$$

The incertitude $\Delta_t^{(\xi)}$ is then given by the difference of two trajectories $\xi_t$ and $\xi_t'$ originated by nearby initial conditions and evolves according the stochastic map

$$\Delta_{t+1}^{(\xi)} = L\,\Delta_t^{(\xi)} + \epsilon\,W_t \tag{12}$$

where $W_t = w_t' - w_t$. For short times $t \ll \lambda_0^{-1}\,|\ln\delta_0|$ one cannot consider the 'noises' $w_t$ and $w_t'$ as uncorrelated so that the incertitude on the driver grows exponentially under the influence on the deterministic chaos given by the feedback, $|\Delta_t^{(\xi)}| \sim |\Delta_t| \sim \exp(\lambda t)$, with $\lambda = \lambda_0 + O(\epsilon)$. For long times $t \gg \lambda_0^{-1}\,|\ln\delta_0|$, the random variables are practically uncorrelated so that their difference $W_t$ still acts as a noisy term. As a consequence the growth of the incertitude is diffusive, since the formal solution of (12) is

$$\Delta_t^{(\xi)} = L^t \left( \delta_0 + \epsilon \sum_{\tau=0}^{t} L^{-\tau}\,W(\tau) \right) \tag{13}$$

and noting that $L^t$ is a unitary transformation, from (13) one can derive the bound

$$|\Delta_t^{(\xi)}| \le \epsilon \left| \sum_{\tau=0}^{t} W_\tau \right| \sim \epsilon\,t^{1/2} \tag{14}$$

where we have used the estimate $|\sum_{\tau=0}^{t} W_\tau| \sim t^{1/2}$ given by standard arguments borrowed from the central limit theorem. In conclusion, for our model maps (10a) and (10b), the predictability time on the driver diverges like

$$T_p^{(\xi)} \sim \epsilon^{-2} \tag{15}$$

although there is a regime of strong chaos since the total Lyapunov exponent $\lambda = \lambda_0 + O(\epsilon)$ does not vanish with the strength of the feedback $\epsilon$.

It is important to stress that the particular power of the diffusive law in a realistic model can be different from that of a random walk, since the deterministic chaos of the feedback could be better represented by random variables with appropriate correlations. The qualitative behavior exhibited by the stochastic model for the incertitude growth (exponential followed by a power law) can be tested in a direct numerical simulation of the coupled maps (10a) and (10b), where we choose the linear vector function for the feedback,

$$h(\eta) = (\eta, \eta) \tag{16a}$$

and the logistic map at the Ulam point for the driving system,

$$G(\eta) = 4\,\eta\,(1 - \eta) \tag{16b}$$

with Lyapunov exponent $\lambda_0 = \ln 2$. Fig 1 shows the behaviors of the incertitude $|\Delta^{(\xi)}|$ for the driver, at starting with an error $\delta_0$ on the initial condition $\eta_0$ and no error on the driver. At the beginning both $|\Delta^{(\eta)}|$ and $|\Delta^{(\xi)}|$ grows exponentially. However, the phase space available to the variable $\eta$ is finite, so that $|\Delta^{(\eta)}|$ is bounded by a maximum value $\Delta_M \sim O(1)$. It will be attained at the time $t = t^* \sim \lambda_0^{-1} \ln(\Delta_M/\delta_0)$. when the incertitude on the driver system is $|\Delta^{(\xi)}| \sim \epsilon \Delta_M$, and so much lower than the threshold. At larger times $t > t^*$, the incertitude on the driven system remains practically constant and $|\Delta^{(\xi)}|$ increases with a diffusive law of type (14) according to the mechanism described by the stochastic map (12).

We have also studied a more realistic model of two coupled standard maps in action-angle variables $I$ and $\theta$,

$$I_{t+1}^{(1)} = I_t^{(1)} - \epsilon \sin(\theta_t^{(1)} + \theta_t^{(2)})$$
$$\theta_{t+1}^{(1)} = \theta_t^{(1)} + I_{t+1}^{(1)}$$
$$I_{t+1}^{(2)} = I_t^{(2)} - K \sin(\theta_t^{(1)} + \theta_t^{(2)})$$
$$\theta_{t+1}^{(2)} = \theta_t^{(2)} + I_{t+1}^{(2)}$$
(17)

where $K \gg \epsilon$ is a control parameter of order unity, such that the system $(I^{(2)}, \theta^{(2)})$ is chaotic when $\epsilon = 0$. We do not discuss in details the results for these coupled maps, since they are qualitatively similar to those obtained for the simplified model (10).

We now consider an application to a physical phenomenon, the motion of an ensemble of point vortices in a fluid. It is a classical problem in fluid mechanics, formally similar to the planetary motion in gravitational field. Both the systems are Hamiltonian with long range interactions. The main qualitative differences are that the Hamiltonian for point vortices does not contain a kinetic term and the motion is confined on the two-dimensional plane. The phase space for a collection of $N$ vortices has thus $2N$ dimensions, related to the physical coordinates.

Dynamical properties of point vortex systems have been studied by several authors interested in their chaotic motion and connection with two dimensional turbulence (see [4] for a review). The Hamiltonian theory for vortex motion inside a bounded domain was developed many years ago [5] for several boundaries. We are here interested in the motion in the unitary disk $D$ for which the Hamiltonian takes the form

$$H = -\frac{1}{4\pi} \sum_{i>j} \Gamma_i \Gamma_j \log\left[\frac{r_i^2 + r_j^2 - 2r_i r_j cos\theta_{ij}}{1 + r_i^2 r_j^2 - 2r_i r_j cos\theta_{ij}}\right] + \frac{1}{4\pi} \sum_{i=1}^{N} \Gamma_i^2 \log(1 - r_i^2) \quad (18)$$

where $\Gamma_i$ represent the circulation of the i-th vortex of coordinates $\mathbf{x_i} = (x_i = r_i cos\theta_i, y_i = r_i sin\theta_i)$ and $\theta_{ij} = \theta_i - \theta_j$. The canonical conjugated variable are the scaled coordinates $(\Gamma_i x_i, y_i)$ and the phase space is thus $N$ times the configuration space $D$. The Hamiltonian (18) is invariant under rotations in the configuration space, thus the angular momentum is a second conserved quantity

$$L^2 = \sum_{i=1}^{N} \Gamma_i(x_i^2 + y_i^2) \quad (19)$$

By general results of Hamiltonian mechanics, a system of two point vortices is always integrable, but we should expect chaotic motion for $N > 2$ vortices.

In the following we will consider $N = 3$ vortices, two of them carrying fixed circulation $\Gamma_1 = \Gamma_2 = 1$ and representing the driver, that without feedback is integrable. The third vortex, of circulation $\Gamma_3 = \epsilon$ represents the driven system which now makes the total system chaotic. In the limit $\epsilon \to 0$ the third vortex becomes a passive particle (it is passively transported by the flow generated by the two unit vortices) and does not influence the integrable motion of the two vortices as for the three-body restricted problem in celestial mechanics. The restricted system is still chaotic, but the incertitude is confined to the passive tracer, while the motion of the two vortices is, in general, quasi-periodic. This limit is one of the simplest example of chaotic advection in two-dimensional flow and it will be studied in detail in another paper [6].

For our particular problem of three vortices the Hamiltonian can be rewritten in the following standard perturbation form

$$H = H_0(\mathbf{x}_1, \mathbf{x}_2) + \epsilon H_1(\mathbf{x}_1, \mathbf{x}_2, \mathbf{x}_3) + \epsilon^2 H_2(\mathbf{x}_3) \qquad (20)$$

The first term $H_0$ describes the dynamics of the two unit vortices (and leads to integrable motion for $\epsilon = 0$); the term $H_1$ represents the interaction with the small vortex of circulation $\epsilon$ and the last term is due to the interaction of the third vortex with its own image. The $O(\epsilon)$ term is thus the perturbation to the integrable system $H_0$ and we are reduced to the general framework described above if we identify $\xi = (\mathbf{x}_1, \mathbf{x}_2)$ and $\eta = \mathbf{x}_3$. The only difference is that now the $\eta$ dynamics is not chaotic by itself, but chaoticity is induced by the interaction with the integrable system $\xi$.

We now describe a typical simulation of error growth in the point vortex model which reproduces the effects obtained with the coupled maps model. We fix the value of the coupling constant (circulation of the third vortex) $\epsilon = 10^{-6}$ and the initial condition for the vortex positions are chosen in order to obtain chaotic motion with a global Lyapunov exponent $\lambda \sim 0.041$. The initial incertitude on the coordinates of the small vortex is $\Delta^{(\eta)}(0) = 10^{-3}$ while we suppose to know the initial position of the two big vortices with a precision of $\Delta^{(\xi)}(0) = 10^{-8}$. The saturation value for the incertitude is proportional to the disk radius, here $\Delta_M \sim 1$.

We let the system evolve according to the Hamiltonian dynamics (18) for quite long time and we computed, at each time, the maximum value reached by the incertitude (we used the maximum because in this system incertitude show strong oscillations: this is a memory of the quasi-periodic behavior for $\epsilon = 0$). This represent the worst situation for making predictions. The upper scatter plot in fig 2 shows the time evolution of the incertitude for the driven system $\Delta^{(\eta)}(t)$. We can recognize a short ($t < 100$) exponential growth until the nonlinear effect becomes important. At large time ($t > t^* \sim 300$) the incertitude saturates to its maximum value $\Delta_M$. The lower scatter plot represents the incertitude for the driver system of two vortices, $\Delta^{(\xi)}(t)$. We can easily recognized the two expected limiting behavior represented by the two lines. For small times, the error grows exponentially, $\Delta^{(\xi)}(t) \sim e^{\gamma t}$ where $\gamma \sim 0.064$ is close to the global Lyapunov exponent. For long time, the power law behavior is recovered, $\Delta^{(\xi)}(t) \sim \epsilon t^{1/\alpha}$ with $\alpha^{-1} \sim 0.88$.

In conclusion we must stress that all our results can generalized in a straightforward way to a weakly chaotic driver with a maximum Lyapunov exponent $\lambda_d \ll \lambda_0$. In fact, the driver might be either conservative or dissipative. The important point is that the dynamics of the driver has a much longer characteristic time than the driven system so that, for an observer interested in the predictability problem, the two systems can be practically decoupled. The Lyapunov analysis, although mathematically correct, does not capture the physically relevant features of the phenomenon, and the exponential dependence on initial conditions does not affect the possibility of forecasting the future of the driver on very long time scale. This is still true in systems with many different time scales instead of only two ones, as fully developed turbulence, where the inverse Lyapunov exponent is not related to the predictability time on the large length scale motion.

Acknowledgements

We are grateful for financial support to I.N.F.N. through *Iniziativa Specifica FI3*. GB thanks the "Istituto di Cosmogeofisica del CNR", Torino, for hospitality.

**Figure captions**

Fig 1 Growth of the incertitude $|\Delta^{(\xi)}|$ of the driver system in the coupled maps (10a) and (10b) as a function of time $t$, where the rotation angle $\theta = 0.82099$, the feedback strength $\epsilon = 10^{-5}$, and the error on the initial condition of the driven system (10b) $\delta_0 = 10^{-10}$. Dashed line: exponential regime $\Delta^{(\xi)}(t) = \epsilon\,\delta_0\,\exp(\lambda_0\,t)$ where $\lambda_0 = \ln 2$. Full line: $\Delta^{(\xi)}(t) = \epsilon\,t^{1/2}$.

Fig. 2 Incertitude growth for the point vortex model. Cross: maximum of the incertitude $\Delta^{(\eta)}(t)$ on the third vortex. Diamond: maximum error $\Delta^{(\xi)}$ on the driving system of two vortices. Dotted line: exponential regime $\Delta^{(\xi)}(t) \sim \exp(\gamma\,t)$ with $\gamma = 0.064$. Dashed line: power law regime $\Delta^{(\xi)} \sim \epsilon\,t^{1/\alpha}$ with $\alpha^{-1} = 0.88$.

REFERENCES

[1] E. N. Lorenz, J. Atmos. Sci. **20**, 130 (1963)
[2] G. Benettin, L. Galgani and J. M. Strelcyn Phys. Rev. **A14**, 2338 (1976)
[3] G. J. Sussmann and J. Wisdom, Nature **257**, 56 (1992)
[4] H. Aref, Ann. Rev. Fluid Mech. **15**, 345 (1983)
[5] C.C. Lin, Proc. Natl. Acad. Sci. U.S. **27**, 570 (1941)
[6] G. Boffetta, A. Celani, P. Franzese and L. Zannetti, in preparation (1995)

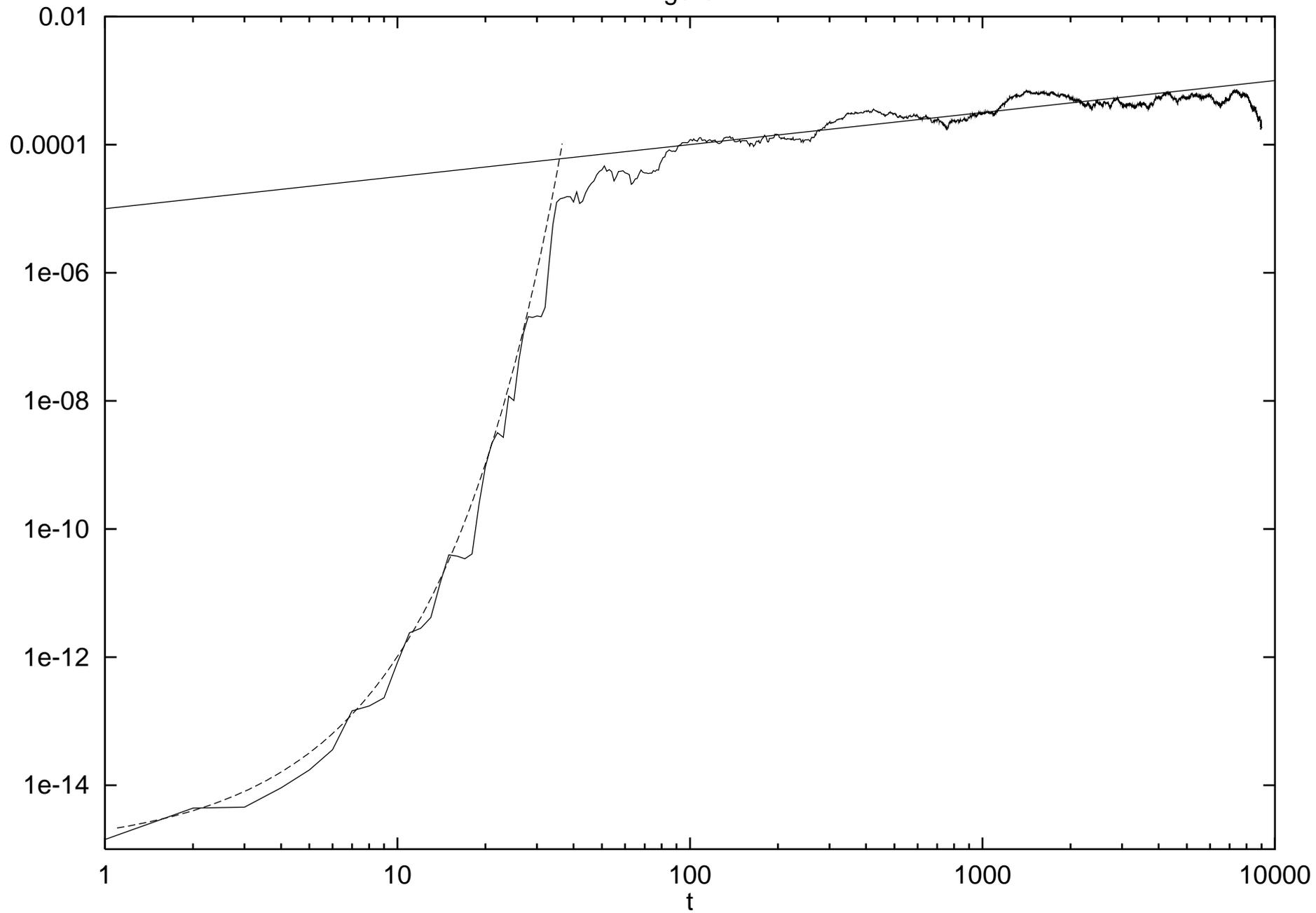

Figure 1

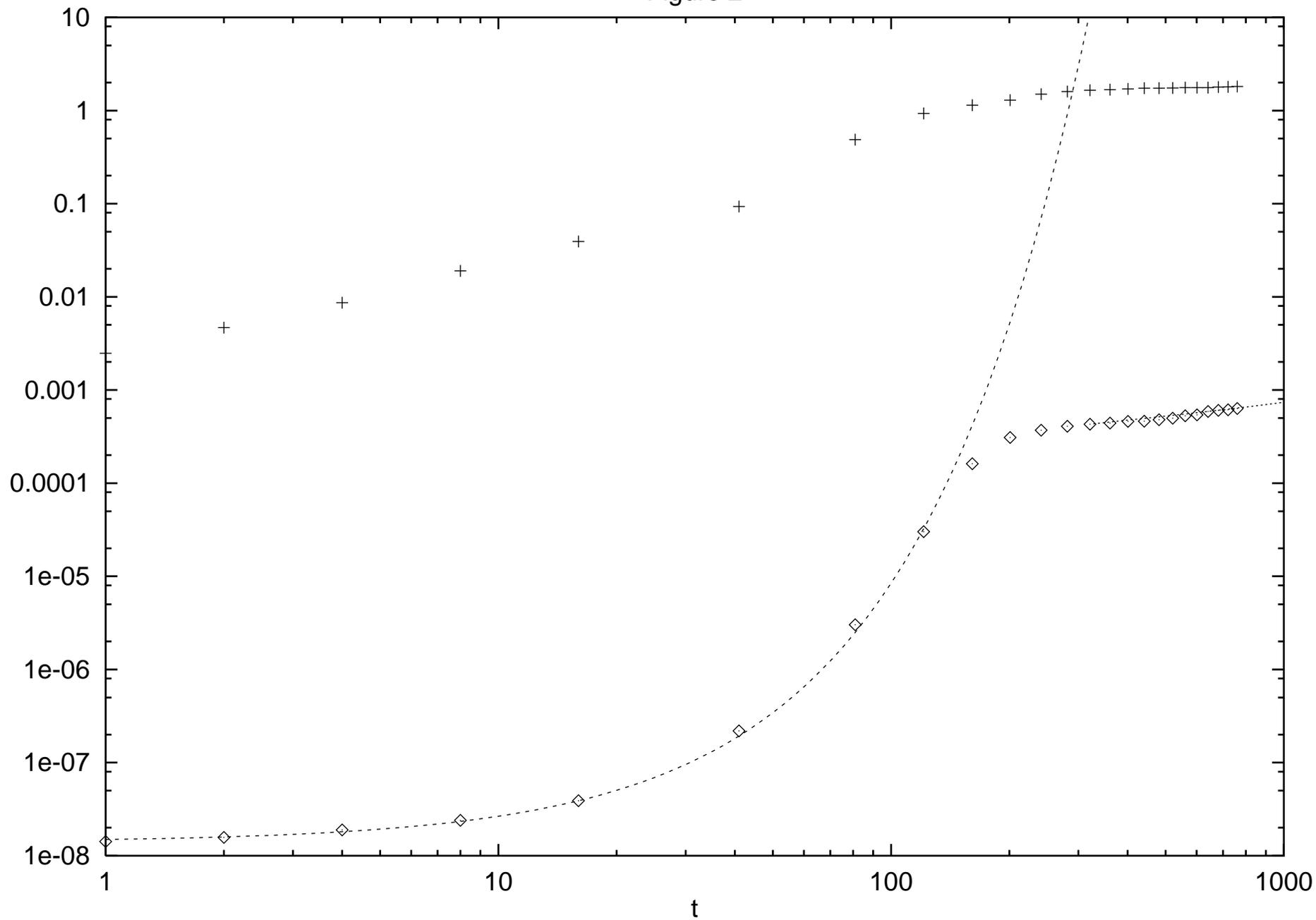

Figure 2